# GoArray: highly dynamic and efficient microarray probe design

Sébastien Rimour[1,*], David Hill[1], Cécile Militon[2], Pierre Peyret[2]

[1] LIMOS UMR CNRS 6158 – Blaise Pascal University –
Clermont-Ferrand II. BP 10125.
63177 Aubiere Cedex, France.

[2] LBP UMR CNRS 6023 – Institut Universitaire de Technologie,
Clermont-Ferrand I University, France.

---

[*] To whom correspondence should be addressed.

**ABSTRACT**
**Motivation:** The use of oligonucleotide microarray technology requires a very detailed attention for the design of specific probes spotted on the solid phase. These problems are far from being commonplace since they refer to complex physicochemical constraints. Whereas there are more and more publicly available programs for microarray oligonucleotide design, most of them use the same algorithm or criteria to design oligos, with only little variation.
**Results:** We show that classical approaches used in oligo design software may be inefficient under certain experimental condition, especially when dealing with complex target mixture. Indeed, our biological model is a human obligate parasite, the microsporidia *Encephalitozoon cuniculi*. Targets that are extracted from biological sample are composed of a mixture of *Encephalitozoon cuniculi* transcripts and host cell transcripts. We propose a new approach to design oligonucleotides which combines good specificity with a potentially high sensitivity. This approach is original in the biological point of view as far as in the algorithmic point of view. We also present experimental validation of this new strategy by comparing results obtained with standard oligos and with our composite oligos. A specific E. cuniculi microarray will overtake the difficulty to discriminate the parasite mRNAs from the host cell mRNAs demonstrating the power of the microarray approach to elucidate the lifestyle of an intracellular pathogen using mix mRNA
**Availability:** Our method is implemented in GoArray software, available at http://www.isima.fr/bioinfo/goarray/
**Contact:** rimour@isima.fr

**INTRODUCTION**

The automation of the Sanger sequencing method coupled to bioinformatics tools development lead to the rapid identification of the CDSs (Coding DNA Sequences) catalogue of a considered organism. This genomic annotation process is applied since several years on hundreds of organisms by large EST (Expressed Sequence Tag) or complete genome projects (Bernal *et al.*, 2001).

Thanks to the increasing number of completely sequenced genomes, the understanding of complex biological processes can benefit of global approaches. Among the recent post-genomic developments, DNA microarray represents probably the most powerful tool. The DNA chip technology has been first developed to evaluate simultaneously the expression levels of all the genes of an organism (DeRisi *et al.*, 1997; Lipshutz *et al.*, 1999; Lockhart *et al.*, 1996; Schena *et al.*, 1995). Other applications different from gene expression monitoring have been applied using DNA chips (specific mutation detections, microbial identification, comparative genomic hybridization, Chromatin Immunoprecipitation chips). The general concept of this nanotechnology development is based on the capacity to immobilize tens of thousand of specific probes on a small solid surface. More than 400 000 sequences representing up to 13 000 genes and expressed sequence tags can be deposited on a surface of 1.6 $cm^2$ (Lockhart and Barlow, 2001). A highly specific recognition step between the probes and the labeled targets (mRNA) is obtained via complementary base pairing during nucleic acids hybridization process. For the production of arrays, DNA can either be directly synthesized on the solid support (Lipshutz *et al.*, 1999) or can be deposited in a pre-synthesized form onto the surface with pins or ink jets printers (Hughes *et al.*, 2001). The spotted DNA corresponds to genomic DNA, cDNAs, PCR products or oligonucleotides.

DNA oligonucleotide microarrays have become more popular compared to cDNA microarrays for gene profiling and are now preferentially set up. Construction of cDNA microarrays presents a number of difficulties, largely related to costs associated with clones validation, quality controls and propagation. Other limitations of cDNA microarrays are observed because of cross-hybridization with homologous genes or alternative splice variants. On the other hand, several reports highlight the excellent sensitivity and specificity of oligonucleotide microarrays and the easiness to obtain these arrays that only require sequence data (Hughes *et al.*, 2001; Kane *et al.*, 2000; Relogio *et al.*, 2002; Wang and Seed, 2003). For example a 60mer oligonucleotides microarray can reach a sensitivity level close to one copy of mRNA per human cell and a single oligonucleotide per gene is sufficient to monitor gene expression (Hughes *et al.*, 2001).

Probes design for microarray is not a trivial computational task. Parameters such as probe length, number of oligos per CDS, maximal distance from the 3' end (to avoid the bias toward the 3' end generated by abortive reverse transcription when probes are labeled using oligodT to anchor the reaction), melting temperature range, threshold to reject secondary structures (to avoid interference during hybridization step), prohibited sequences and of course specificity have to be considered to obtain an efficient design. The cross-hybridization is the major point that limits the determination of specific probes. During classical hybridization experiments, cross-hybridization is limited by increasing the stringency of hybridizations conditions. However, it is not possible to apply the most stringent conditions for all the genes of an organism at the same time. In fact, cross-hybridization of a CDS to other sequences may result from a single continuous alignment exceeding cross-hybridization parameters or from a discontinuous series of short similarity regions (within the same subject) appearing as several alignments which all together exceed the cross hybridization parameters. In the yeast *Saccharomyces cerevisiae* 253 CDS (4,5% of the total CDSs) failed to be represented by a "unique" oligonucleotide probe (Talla *et al.*,2003). Even with an improved version of OligoArray (2.0) where the computation of the specificity is based on the thermodynamics of the hybridization, 7% of the *Arabidopsis thaliana* CDSs stay alone without the possibility to design a specific probe (Rouillard *et al.*, 2003). The percentage will increase with the complexity of the considered biological model.

Concerning the specific probe design for a human obligate parasite like our biological model, the microsporidia *Encephalitozoon cuniculi*, the situation will be more dramatic. By conventional probe design we were able to determine specific 50-mers oligonucleotides for only approximately 40% of the CDSs. These results prompted us to develop a highly dynamic probe design (GoArray) that lead to the determination of specific oligonucleotide for almost all the CDS except of course for perfect gene duplication.

The remainder of this paper is organized as follows. In the Methods section, after presenting a short review of existing software for microarray oligonucleotide design, we show the limits encountered, particularly when dealing with complex biological models, and we propose a new oligo design approach which enables to bypass the previously cited limits. Implementation section deals with our computational choices for the implementation of this new approach. Then, we present the experimental verification before discussing the observed results.

## METHODS

### Related Works

Whereas there are more and more publicly available programs for microarray oligonucleotide design, most of them use the same algorithm or criteria to design oligos, with only little variation (Chen and Sharp, 2002; Li and Stormo, 2001; Nielsen *et al.*, 2003; Raddatz *et al.*, 2001; Rahmann, 2003; Reymond *et al.*, 2004; Rouillard *et al.*, 2002; Rouillard *et al.*, 2003; Talla *et al.*, 2003; Wang and Seed, 2003). They primarily differ by the criteria chosen to select the oligonucleotides: some will take into account the possibility for an oligonucleotide of having a stable secondary structure under certain conditions, others will make it possible to exclude certain sequence patterns determined by the user, such as the repetition of a single base. None of them gathers the whole set of criteria.

From a data-processing point of view, one can separate software in two groups of software: the client/server type and the autonomous software type. In the case of an application of the client/server type, the user employs a small software client to send the data to a server, generally located in the laboratory which developed the design software. This laboratory is carrying out the calculation of the oligos and then is sending the results to the customer. The advantages of this type of applications are mainly that the final user does not need to have large local computing facilities at his disposal and that the software client is often very easy to install and to use over the Internet.

The major drawback of this type of remote application is that the database used to test the specificity of the oligos must be present on the server, which limits the possibilities. Indeed, we generally finds on servers only the principal genomes studied in biology, those of the model organisms. If the user carries out DNA Chips experiments on little studied organisms, he will not be able to use such software or will have to contact the administrator of the server to add a database. In addition, the achievement of calculation is dependent on the server availability and the confidentiality of the data is not guaranteed though some scientists are currently explicitly working on this point (Kurata *et al.*, 2003).

However when the server software is easily accessible, it is possible to install it on a dedicated machine, we return then to the behaviour of autonomous software. In the case of autonomous software, the user completely controls the configuration of the parameters of the oligonucleotides design and is not dependent on a distant machine. However, the applications of this type are more difficult to install, require a certain computing power and sometimes need other software to function (BLAST (Altschul *et al., 1997)*; Mfold (SantaLucia, 1998; Zuker *et al.*, 1999).

*Client Server software.* OligoWiz (Nielsen *et al.*, 2003) is composed of a free Java client (data input and display of results) and a Perl server (under Unix) hosted by CBS (Center for Biological Sequence analysis, Denmark: http://www.cbs.dtu.dk/services/ OligoWiz/). The oligonucleotide searching method is the following: for each potential oligonucleotide, the program takes into account 5 parameters and for each one of these parameters, it calculates a score for the oligonucleotide. It makes then a balanced sum (by user-defined coefficients) of the scores and returns the oligonucleotide having obtained the best score. The originality of this software is that the user can also graphically visualize the scoring functions of all the potential oligonucléotides and manually select some of them. The criteria taken into account for the search are: specificity, melting temperature (Tm), position within transcript, complexity of the sequence and its composition in bases other than ATCG.

Another client-server software (ROSO) has been recently proposed by the French Biopole of Lyon (Reymond *et al.*, 2004). Accessible via a Web site, the software is written in C and uses BLAST to compute the specificity (http://pbil.univ-lyon1.fr/roso/). It is possible to obtain the source and executable code on demand. The search algorithm is composed of 5 steps :

1) Filtering of the sequences provided in input (elimination of identical genes, of the repetitions of bases...).
2) Search of potential cross hybridizations using BLAST.
3) Elimination of oligonucleotides having a stable secondary structure.
4) Calculation of the Tm of each candidate oligonucleotide and selection of a set of oligonucleotides (at least one by gene) which minimizes the variability of the Tm.
5) Selection of the final set of oligonucleotides on 4 criteria (composition in GC, first and last bases, repetitions, free energy).

*Autonomous Software.* OligoArray (Rouillard et al., 2002) is written in Java, freely distributed and its source code is available under the GPL (General Public Licence). BLAST is necessary for its functioning along with the MFold software with a mandatory access over Internet for the computation of the oligonucleotide secondary structures on the corresponding server. This could have classified OligoArray in the category of client/server software but most of the computation is done locally. In addition it is possible to install MFold locally but then the software is less reliable. The criteria taken into account for the search for oligonucleotides are: specificity, position in transcript, melting temperature, secondary structure of the oligonucleotide, and presence of specific patterns in the sequence of the oligonucleotide. The strategy adopted by OligoArray is as follows: given a sequence provided in input, it extracts a sub-sequence made up of the 'n' last bases (N being the length desired for the oligos) and checks the set of the criteria cited above. If this sub-sequence meets all the criteria, it will constitute a satisfactory oligonucleotide, and the program passes to following gene. In the contrary case, the program carries out a shift of 10 bases towards extremity 5' of the CDS, extracts again a sub-sequence and checks the criteria. The process continues until a satisfactory oligonucleotide is found. If no oligonucleotide is found, it makes a new screening of the sequence with a less strict specificity criterion. If again no oligonucleotide is found, the program returns the oligonucleotide presenting the less potential cross hybridization. For the whole set of criteria considered, the program returns a binary response: if the oligonucleotide meets the criterion, it is kept, if not it is rejected. A returned oligonucleotide is the sequence nearest to the 3' extremity which meets all the other criteria. It is also possible to define a limit to stop the searching beyond a certain distance to the 3' extremity (http://berry.engin.umich.edu/oligoarray/). OligoArray 2.0 (Rouillard et al., 2003) is an evolution of the software with a slightly diffrent approach using thermodynamics to design oligonucleotides (see: http://berry.engin.umich.edu/oligoarray2/).

Another software named ProbeSelect retained our attention though the algorithm used is quite complex. ProbeSelect is written in C++ under Sun Solaris (Li and Stormo, 2001) and the approach to find oligonucleotides is centered on the specificity of sequences. Its functionning requires seven steps :
1) Construction of a "suffix array" for the set of the coding sequences issued from the genome of the considered organism.
2) Construction of a "landscape" for each gene provided in input
3) These structures are used to determine a list (10 to 20) of candidate oligonucleotides for each gene (sequences which minimize the sum of the frequencies of theirs sub-words in the whole studied genome)
4) For each oligonucleotide candidate, search of potential cross hybridizations
5) Localization of these crossed hybridizations within genes
6) Calculation of free energy and of the Tm for each hybridization
7) Selection of the oligonucleotides which have the most stable hybridization with their target and which allow a good discrimination of the others potential targets.

The considered criteria to obtain a satisfactory oligonucleotide are thus not evaluated separately, but are nested in the algorithm. There is a potential cross hybridization if an identical sequence (other than the target gene) to the oligonucleotide with a certain number of mismatches authorized is found in the genome: 4 mismatches for the short oligonucleotides from 20 to 25 bases ; 10 mismatches for the long oligonucleotides of 50 bases and 20 mismatches for the oligonucleotides of 70 bases.

Some autonomous oligonucleotides design software are written in Perl or are based on a set of Perl scripts. They are usually much slower than from the C++-written ones. OligoPicker (Wang and Seed, 2003) is a Perl program which also uses a traditional approach. The criteria taken into account are specificity (use of BLAST), melting temperature and position in the transcripts (http://pga.mgh.harvard.edu/oligopicker/). ProMide (Rahmann, 2003) is a set of Perl scripts, with a C program for main computation, that use longest common factor as a specificity measure for the oligonucleotides. It uses complex data structures such as ("enhanced suffix array") and some statistical properties of sequences (http://oligos.molgen.mpg.de/). Oliz (Chen and Sharp, 2002) is also a set of Perl scripts using a classical approach (using BLAST for specificity testing). It however requires additional software to function (cap3, clustalw, EMBOSS prima) plus a database in the UniGene format. The originality of the method used comes from the fact that the oligonucleotides are searched in area 3' UTR (untranslated region) of the mRNAs, a very specific area where sequences are largely available through EST (Exprssed Sequence Tag) projects (http://www.utmem.edu/pharmacology/otherlinks/oliz.html).

**Limits of existing software**

*The Specificity Problem.* Concerning the design of probe sets for microarrays, the specificity of oligonucleotides is one of the most important points, among all criteria that probes must satisfy. The optimum probe for a gene should hybridize only with the transcripts of this gene (under hybridization conditions) and not with other transcripts present in the hybridization pool. To select such a probe, one must know in what condition an oligonucleotide could hybridize with a non-target gene, especially in terms of sequence homology. Although few biological studies were made to answer this question, a very complete one has been achieved on 50mers oligonucleotides (Kane *et al.*, 2000). They show that a probe must satisfy two conditions to be specific:
(1) The oligonucleotide sequence must not have more than 75% of similarity (among all sequence) with a non-target sequence present in the hybridization pool.
(2) The oligonucleotide sequence must not include a stretch of identical sequence > to 15 contiguous bases.
Many probe design software use Kane's criteria to verify the specificity of oligonucleotide (OligoArray, OligoWiz), and they extend it for oligonucleotides of any length. ProbeSelect uses the number of mismatches between the oligonucleotide and the non-target sequence. They may cross-hybridize if there are four or fewer mismatches for short oligos (around 20mers), 10 or fewer mismatches for 50mers, and 20 or fewer for 70mers.
In our study, we follow Kane's criteria and consider that a probe may show cross-hybridization with a non-target sequence when it does not satisfy condition (1) and (2). In a rather obvious way, the longer the oligonucleotide is, the easier is condition (1) satisfied. However, it is more complex for condition (2): as the length increase, the probability of finding a stretch of 15 identical bases may become more important.

*Limits when dealing with a complex biological model.* The microarrays we produce are designed to elucidate the adaptation mechanisms involved in a parasitic life style. Our biological model is the emerging opportunistic parasite *Encephalitozoon cuniculi*. Recently, the genome of this pathogen, considered as the smallest eukaryotic genome with a size of only 2.9 Mb, has been completely sequenced by our research team and the French Genoscope (Katinka *et al.*, 2001). This parasite represents an increasing danger in human health because of the increase in the number of immuno-depressed people (following diseases like AIDS, chemotherapies...). Regarding microarray experiments, the originality of our study is that targets which are extracted from biological samples are composed of a mixture of *Encephalitozoon cuniculi* transcripts and host cell transcripts (human cell in our case), we can't separate them. A specific *E. cuniculi* microarray will overtake the difficulty to discriminate the parasite mRNAs from the host cell mRNAs demonstrating the power of the microarray approach to elucidate the lifestyle of an intracellular pathogen using mix mRNA. Thus, the oligounucleotides designed for the genes of the parasite must be specific not only regarding the whole parasite CDSs, but also regarding human genome (as human transcripts are present in the target mixture). This induces a supplementary difficulty in the design of specific probes by increasing the complexity of the database.
When we try to design 50mers probes for our biological model with classical software (using Kane's criteria), we find only a few specific ones. Most of the oligonucleotides show cross-hybridization with human genes, because they don't satisfy condition (2) of Kane's criteria. If we reduce the length of the oligonucleotides, we may find more oligonucleotides satisfying this condition but the risk of having more than 75% of similarity with a non-target sequence increases. In the next part, we perform in silico tests to explore the influence of oligonucleotide length on specificity.

*In Silico Tests.* We performed in silico tests on two databases: the relatively small genome of the yeast *Saccharomyces cerevisiae* and our model *Encephalitozoon cuniculi*. In the second case, when we want to design an oligonucleotide, we must check its specificity against a database that consists of the union of *Encephalitozoon cuniculi* genome and human genome (Unigene database). For each organism, we randomly chose 100 CDSs and for each CDS, we checked the specificity of a set of oligonucleotides extracted from these CDSs (non-overlapping oligos). We report how many times an oligonucleotide "matches" with a sequence present in the database. An oligonucleotide "matches" when it does not satisfy Kane's criteria, thus it is not specific. Hits are found with the BLAST algorithm, and then the whole oligonucleotide sequence is aligned with subject sequence in the database to check Kane's conditions. The total number of matches is considered as a measure of specificity of oligonucleotides of a given length. The results are shown in figure 1.
The same trend is observed for the two organisms: 20-mers oligos have many more matches with the database than other lengths and may not be suitable for oligonucleotide design. However, from 30 mers to 80 mers, we can observe that the specificity is reduced with the increase of the oligonucleotide size. This confirms that for long oligonucleotides, many hits are achieved because they do not satisfy condition (2) of Kane's criteria. Values for *Encephalitozoon cuniculi* are higher and the increase is more marked as the database contains more sequences. These results highlight the risk of cross-hybridisation with long oligo microarrays in a complex biological system. Shorter oligonucleotides (above 20mers) increase specificity but could induce a lower sensitivity. Furthermore, Kane's criteria may be too strict for short oligonucleotides and in the rest of the article, we modify the condition (1) by considering the percentage of similarity only among the BLAST alignment and not among all sequence. That enables us to find specific sequence even shorter (20-25mers) for some CDSs.
In the next part, we present a new method for oligonucleotide design that enables us to solve the problem of specificity in a complex biogical system keeping a high sensitivity with long oligonucleotides.

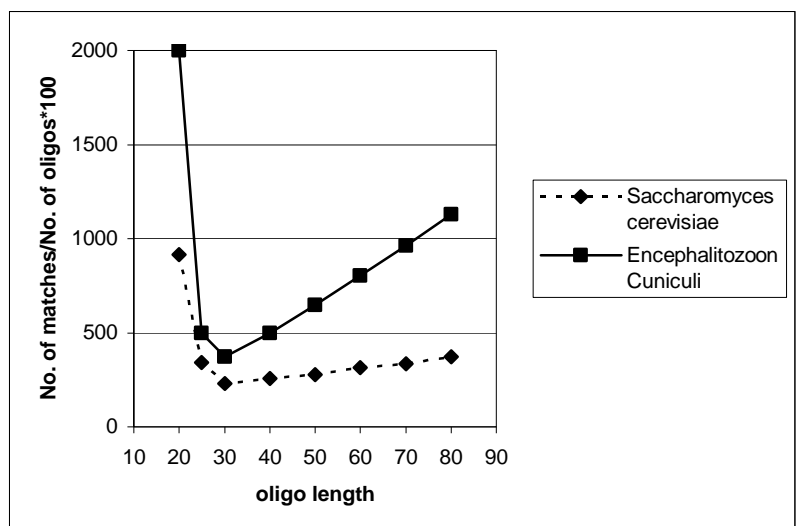

**Fig. 1.** In silico estimation of length-specificity relationship in *Saccharomyces cerevisiae* (yeast) genome and in a complex biological system (*Encephalitozoon cuniculi* + human transcripts). For each organism, a set of oligonucleotides (non-overlapping oligonucleotides from 100 randomly chosen CDS) is tested for specificity against the database. Total number of matches with the database is reported. For a short oligo length, more oligonucleotides are tested than for longer lengths. Thus, the number of matches are normalized to the total number of oligonucleotides tested of a given length.

**A New Approach**

*New kind of oligonucleotide structure.* The "in silico" test shows that in a complex biological system, "long" oligonucleotides are not suitable for good specificity. Specificity increases as the oligonucleotide length decreases and short oligonucleotides between 20 and 30mers seem to be more adapted. However the use of such probes could induce a lower sensitivity. We propose a new approach to design oligonucleotides that combines excellent specificity (even with a complex database) with a potentially high sensitivity.

In our strategy, the oligonucleotide sequence is the concatenation of two sequences that are complementary to their cDNA target but disjoint (figure 2). Thus, the determination of an oligonucleotide is made by searching two "short" sequences (e.g. 25 mers) that are each specific of the CDS (it is easier to find two short specific sequences than one longer). Stable hybridization between the composite probe and the cDNA target induces the formation of a loop. In the oligonucleotide sequence, we also insert a very short randomly linker (3-6 bases) between the two specific sequences to facilitate the formation of the transcript loop. As the oligonucleotide sequence length is still quite long (e.g. 55mers), we keep the advantage of high signal intensity of long oligonucleotides. We also propose a program that enables to design oligonucleotide for microarray according to this new approach.

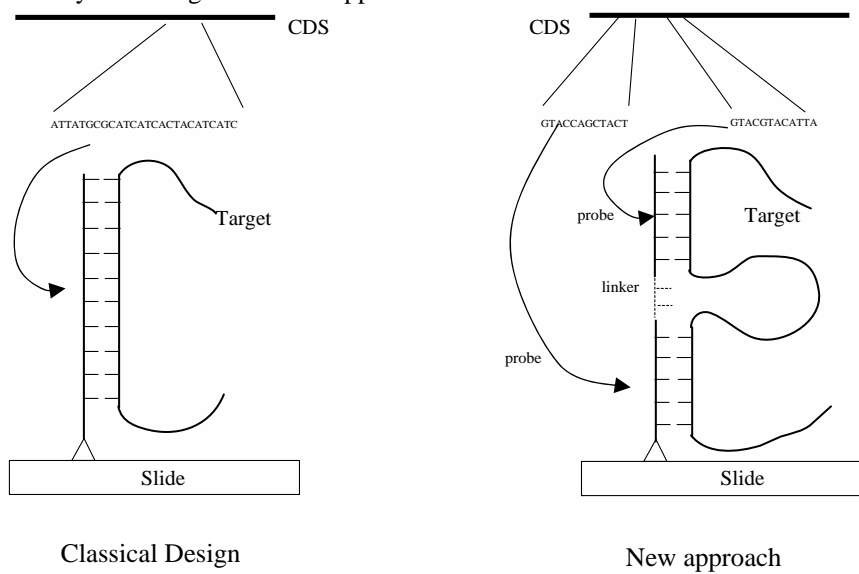

**Fig. 2.** Scheme of the new design of oligonucleotide for microarrays compared to classical design. In the common approach, the oligonucleotide is a specific subsequence of the CDS. In our new approach, the oligonucleotide is composed of two non-contiguous sequences from the CDS. Stable hybridization between the composite probe and the cDNA target induce the formation of a loop. A small linker composed of randomly chosen bases is added between the two sequences.

**IMPLEMENTATION**

For the software analysis, we defined a Platform Independent Model (PIM) for oligonucleotide design compliant with the current MicroArray Gene Expression Object-Model (MAGE-OM) (OMG, 2003) specified by the Object Management Group (OMG). The MAGE-OM is a Unified Modeling Language (UML) model which attempts to define standard objects for gene expression data interchange. Before any programming we have proposed a metamodel of oligonucleotide design, which is MAGE-OM compliant in order to follow state of the art development as advised by the Model Driven Architecture (MDA) (Hill *et al.*, 2002). This rather important choice increases the software quality and potentiality of the algorithm and will now be easily integrated in a microarray software environment. We used this metamodel to implement our oligonucleotide design software: GoArray. This software enables to design specific probe for microarray according to our new approach. It is particularly adapted to complex biological system, when classical softwares are inefficient. The GoArray prototype is written in Java, requires Blast standalone program and MFold software if user wants to check secondary structure of the oligos. The program takes several parameters defined by user via a graphical interface. It requires two input files, one containing all sequences for which we want to design an oligonucleotide, and the second containing the specificity database. This second database must contain all the sequences that may be present in the targets mRNA mixture during the hybridization step. The user chooses the length $l$ of the two subsequences constituting the oligonucleotide, the minimal and maximal loop size $lo_{min}$ and $lo_{max}$, and the size of the linker separating the two subsequences. He also sets the specificity threshold, which is the minimal length of a stretch of identical bases with non-target sequence that may lead to cross-hybridization (default is Kane's criterion: 15 bases). Then, the user can choose if he wants the oligonucleotide to satisfy other classical criteria: melting temperature, absence of stable secondary structure, absence of forbidden sequences. If it is the case, he sets the minimal and maximal melting temperature desired, the maximal melting temperature for secondary structure.

The algorithm takes each input sequence and reads it from 3'end. It tries to find the first specific subsequence by moving a window of length $l$. The specificity of the subsequence is checked by performing the BLAST algorithm on the specificity database. If the subsequence contains a stretch of 15 identical bases with non-target sequence or shows an alignment with 75% or more similarity, it is considered non-specific. Then the program searches the second specific subsequence with the same method, taking into account $lo_{min}$ and $lo_{max}$. When the two specific subsequences have been found, a small linker made of randomly chosen bases is added between the two subsequences. The specificity of the whole sequence is checked in order to verify that the introduction of the linker does not induce supplementary cross-hybridization.

The last step is the verification of additional constraints if needed. Melting temperature of oligonucleotides is computed using the Nearest-Neighbor model (SantaLucia, 1998), with same parameters as OligoArray software. The program calls the MFold software to check if the oligonucleotide has a stable secondary structure, and if it is the case, the sequence is rejected. The last point checked is the presence of forbidden sequences defined by the user (like repetition of bases). From a software development point of view, it's easy to add other constraint as it is facilitated by our object oriented design (Hill *et al.*, 2002). When a sequence is rejected, the windows of the two subsequences are moved along the sequence, until a correct oligonucleotide is found.

The next section will present some experimental verification we achieved with genes from *Encephalitozoon cuniculi*.

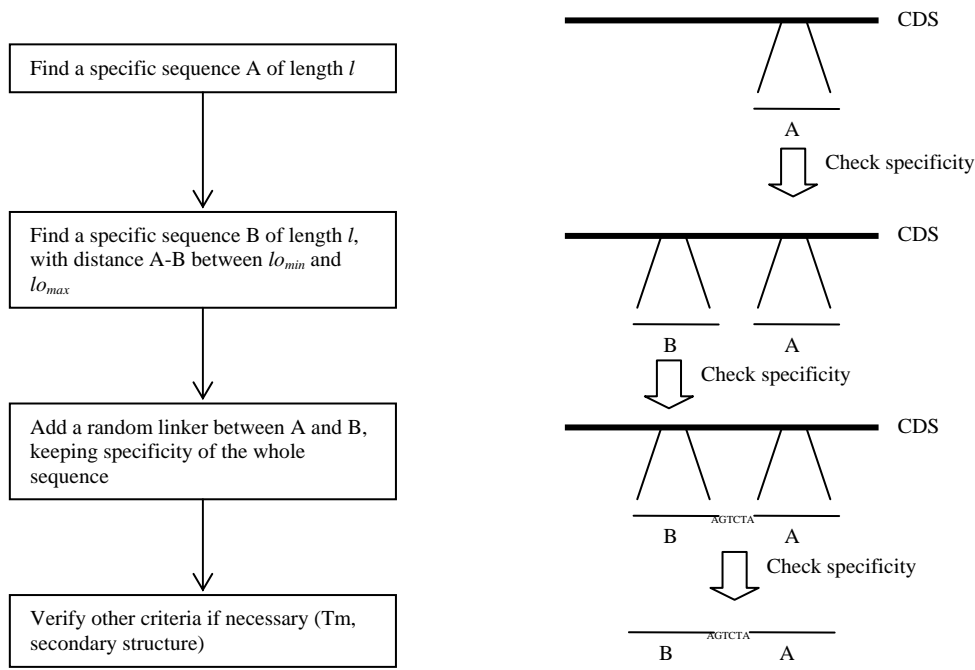

**Fig. 3.** Diagram describing GoArray algorithm. *l* is the length of the two subsequences constituting the oligonucleotide, $lo_{min}$ and $lo_{max}$ are the minimal and maximal loop size.

## RESULTS

### Oligonucleotide design

First we focused our study on two genes of *E. cuniculi*: polar tube protein *ptp1* and cytidylate kinase *kcy* that could induce cross-hybridization with standard oligonucleotide design. We designed a 50mer oligonucleotide for each of these two genes with OligoArray software (Rouillard *et al.*, 2002). Oligonucleotide probe *ptp1* shows a sequence similarity with *kcy* gene (16 identical bases on 17) thus there is a risk of cross-hybridization with *kcy* transcripts depending of the hybridization conditions. Oligonucleotide probe *kcy* shows sequence similarity only with one human gene, and does not cross-hybridize with any *Encephalitozoon cuniculi* gene. The characteristics of these two standard oligonucleotides are summarized in table 1.

**Table 1.** Characteristics of standard 50mer oligonucleotides for two genes of *Encephalitozoon cuniculi*. There is a risk of cross-hybridization between *ptp1* probe and *kcy* transcripts.

| Name | Sequence | Position within transcript | Tm | Possible cross-hybridization |
|---|---|---|---|---|
| *ptp1* | TAGGAACATGCAAGATTGCCGTATTGAAGCACTGCGACGCACCAGGAACA | 896 | 90 | *kcy* |
| *kcy* | AGGAACAACAGGGTATTCCTAGACGGAGAGGACGTGTCGGAGAGCCTCCG | 450 | 91 | gnl\|UG\|Hs#S1731224 |

We also designed oligonucleotide probes for these two genes with our new algorithm GoArray. We calculated 25 oligonucleotides for each gene and design chimeric oligonucleotides (two short specific oligonucleotides linking by 3, 4, 5 or 6 random nucleotides). Variations in chimeric oligonucleotides include: size of the loop produced by the labelled target after hybridization, size of the linker between the two specific sequences, specificity threshold, and size of the two subsequences constituting the oligonucleotide. The specificity threshold is the minimal length of a stretch of identical bases with non-target sequence that may lead to cross-hybridization (15 bases in Kane's criteria, we also tested 16). The oligonucleotides are named regarding their characteristics, for example, the probe PTP1_30_6_16_20 produces a loop size of 30 bases, the linker is 6 bases long, the specificity threshold is fixed at 16 bases and the length of the two subsequences is 20 bases.

### Experimental procedure

For the Microarray production, the oligonucleotides probes (50mers) were synthesized by Eurogentec (http://www.eurogentec.be) with a 5' amino linker modification and the spotting was made by Eurogentec on aldehyde activated slides. The RNA transcripts *ptp1* and *kcy* (genes from *Encephalitozoon cuniculi*) were amplified by PCR reactions using the following primers (The bold part of the oligonucleotides permits the formation of a T7 promoter) as described in the Molecular Cloning third edition (Sambrook and Russell, 2000):

ptp1DT7: **TAATACGACTCACTATAGGTACT**TTGCCCTGATGAAGTTGGA
ptp1R: CAGCAGTGTTGCATGGAGA

kcyDT7: **TAATACGACTCACTATAGGTACT**CAAGATTGCCGTTGATGG
kcyR: TTCGTCGGCTGTTTCCTC

In vitro transcription was conducted using MEGAscript kit from Ambion (http://www.ambion.com/) according to the manufacturer instructions. Then the labeling of the RNA transcripts (3 µg) with Cy3 and Cy5 was conducted using the Amino Allyl cDNA Labeling kit from Ambion (http://www.ambion.com/) according to the manufacturer instructions. Dealing with hybridizations, they were carried out using 25 µl sample (DigEasy buffer from Boehringer 17µl, labelled cDNA 5µl, Salmon sperm DNA 1µg/ml 3 µl) under a supported coverslip at 37°C (water bath) for 16h in a Corning chamber. After hybridization the microarrays were washed two times at room temperature during 5 min with the following solutions (solution1: 0,2X SSC, 0,1% SDS; solution 2: 0,2X SSC). The slides were scanned on the Affymetrix 428 Array scanner to detect Cy3 and Cy5 fluorescence and raw data analysis was carried out using Jaguar from Affymetrix (http://www.mwg-biotech.com/).

### Comparative studies using OligoArray (standard design) and GoArray (new design) for the oligonucleotides design

Figure 3 shows the two images of the microarray obtained after hybrdization with *ptp1* mRNA labelled with Cy3 and *kcy* mRNA labelled with Cy5. PTP1 standard oligonucleotide (OligoArray design) hybridizes with high intensity like some new design oligonucleotides (but not all of them). The same phenomenon is observed on Cy5 image with KCY oligonucleotides.

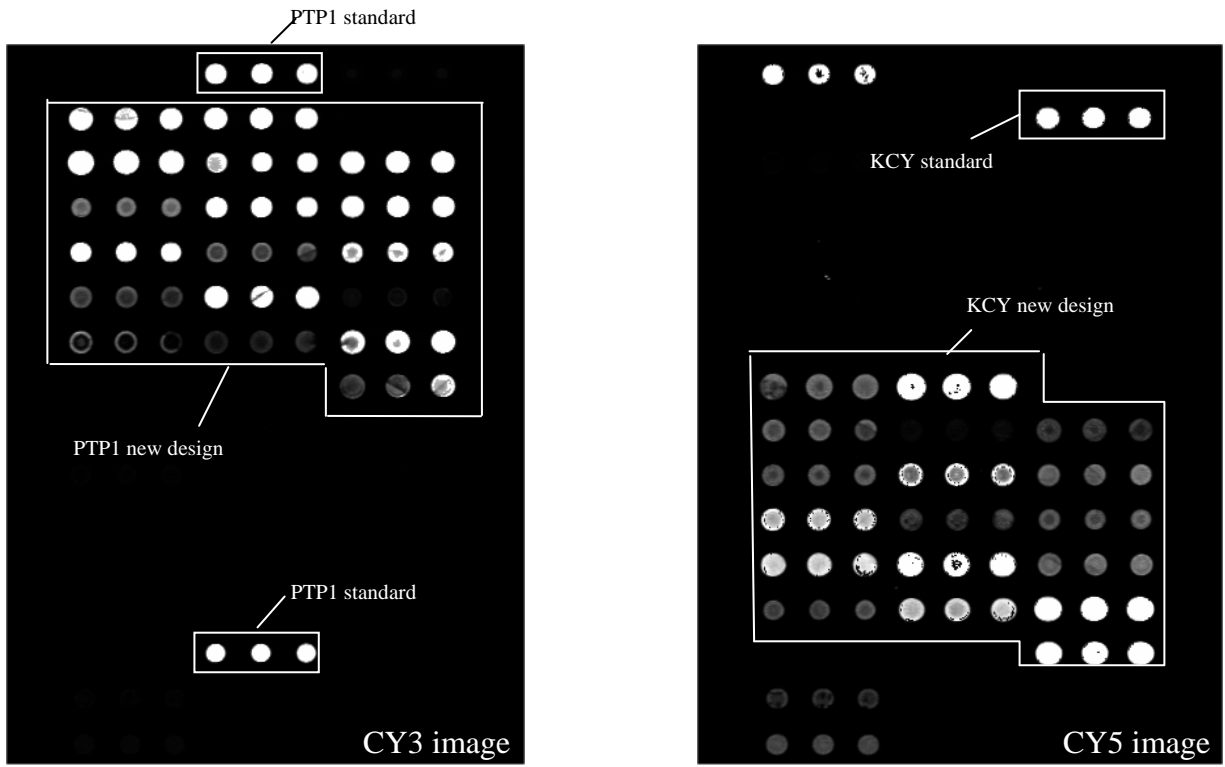

**Fig. 4.** Images of the test microarray. For two genes of *Encephalitozoon cuniculi*, probes were designed with standard method (OligoArray) and with new method (GoArray) and spotted on the array. Slide was hybridized with *ptp1* transcripts labelled with Cy3 and *kcy* transcripts labelled with Cy5.

These observations are confirmed by the measurement of signal intensities with the Jaguar (Affymetrix) image analysis software (table 2). The mean signal of all new design replicates is lower than the standard oligonucleotide, and the variations between new design oligonucleotides with different characteristics are very important. This emphasizes that some oligonucleotides do not hybridize with a high efficiency with their target. So, it is important to determine the best characteristics of the oligonucleotides avoiding cross-hybridizations and giving high sensitivity.

For each gene, we selected the four new design oligonucleotides that gave highest signal intensity. In order to check that the hybridization of the oligonucleotide with its target was actually on the entire length of the probe, we also spotted, for some of the oligonucleotides, the two corresponding 20mers subsequences A and B that constitute the chimeric oligonucleotides, and a third oligonucleotide that is compound of one subsequence of 20mers (A or B) linked with a random sequence of approximately 20mers. Signal intensities of those oligonucleotides are reported in table 3.

**Table 2.** Comparison of signal intensities of standard oligonucleotides and new design oligonucleotides (mean and standard deviation of all replicates)

|  | Signal intensity of standard oligos | | Signal intensity of new design oligos (all) | |
| --- | --- | --- | --- | --- |
|  | Mean | Standard Deviation | Mean | Standard Deviation |
| *ptp1* | 18277 | 1213 | 14989 | 9793 |
| *kcy* | 44231 | 3744 | 13793 | 9638 |

**Table 3.** Signal intensities of new design oligonucleotides with GoArray. Signal intensities of the four oligonucleotides that have highest signal are reported and compared with the corresponding 20mers oligonucleotides subsequences. For example, the oligonucleotide KCY_BACSU_10_4_15_20 is composed by the subsequence A (20mers) + the linker (4mers) + the subsequence B (20mers), then we spotted sequence A alone, sequence B alone, and A + a random sequence (20mers).

| Oligonucleotide new design | Signal intensity | Signal intensity of corresponding 20mers oligo (sequence A) | Signal intensity of corresponding 20mers oligo (sequence B) | Signal intensity of oligo 48mers with seq A + random sequence |
|---|---|---|---|---|
| KCY_BACSU_10_4_15_20 | 32842 | 1355 | 20805 | 5176 |
| KCY_BACSU_10_5_16_20 | 33896 | - | - | - |
| KCY_BACSU_30_6_16_20 | 42011 | 10724 | 1820 | 8052 |
| KCY_BACSU_100_6_16_20 | 33919 | - | - | - |
| PTP1_30_4_16_20 | 31611 | 412 | 6291 | 2146 |
| PTP1_30_5_16_20 | 31683 | - | - | - |
| PTP1_30_6_16_20 | 33499 | 2575 | 11 | 4051 |
| PTP1_40_6_16_20 | 31403 | - | - | - |

For *kcy*, the signal of the new design oligonucleotides (from 33842 to 42011) is just a little bit lower than standard oligonucleotide (44231). The signal of new design oligonucleotides is approximately 72% of standard signal, except one oligonucleotide which is nearly at the same level as for standard oligonucleotide. Concerning *ptp1*, the trend is inverted: the signal of new design oligonucleotides (from 31403 to 33499) is higher than standard oligonucleotide (18277). The signal of corresponding 20mers sequences is much lower and quasi zero for some of them.

These results clearly demonstrate that oligonucleotides designed with our new method (GoArray) can hybridize their target with a high specificity and sensitivity. We demonstrate that the hybridization of our chimerical oligonucleotides is effective all along the two subsequences without a destabilisation induced by the loop formation. We also hybridized microarray with biological sample composed of complex mRNA and proved the efficiency of our design (data not shown).

**New oligonucleotides design with GoArray avoid cross-hybridization**

Hybridization of test microarray described in section 4.2 was repeated several times. We detected a signal intensity of *ptp1* standard oligonucleotide in Cy5 image, which represents cross-hybridization of this oligonucleotide with *kcy* transcript. This result demonstrates that under certain conditions of hybridizations (salt concentration, temperature variation), cross-hybridization may occur. We never observed cross hybridization with our new design. Table 4 shows signals in Cy5 channel for standard *ptp1* probe, and for two new design probes. The latters do not give any signal whereas standard PTP1 shows a signal. Under real condition of hybridization, this signal would be added to the target signal (kcy) and would modify the measure inducing a misinterpretation of the biological proces. Our design enables to avoid cross-hybridization by offering more candidates when searching for highly specific sequences within the CDS.

**Table 4.** Cross-hybridization measurement. Cy5 signal for PTP1 oligonucleotide indicates a cross-hybridization between PTP1 probe and *kcy* transcript. This cross-hybridization has never been observed for new design probes.

|  | Signal intensity | Background intensity |
|---|---|---|
| PTP1 standard | 1361 | 93 |
| PTP1_30_6_16_20 | no signal | 53 |
| PTP1_30_5_16_20 | no signal | 50 |

**Hybridization with complex biological sample**

We also hybridized a microarray with more complex biological sample, under conditions that are closer from real transcriptomic studies of our biological model. Targets were composed of a mixture of *Encephalitozoon cuniculi* transcripts and host cell transcripts (human).
Table 5 reports spot intensities in Cy3 channel for PTP1 standard oligo as well as a selection of chimerical oligos. Although these results are more difficult to interpret than previous ones, they seem to be promising. Chimerical oligos hybridize with targets, signals are even higher than standard oligo. However, signal variation between standard oligo and chimerical oligos as well as signal variation among chimeric oligo is still unclear for us.

**Table 5.** Hybridization with complex biological sample. Signal and background intensities are reported for ptp1 standard oligo as well as for a selection of new design oligos.

|  | Signal intensity | Background intensity |
|---|---|---|
| PTP1 standard | 5126 | 2235 |
| PTP1_10_5_16 | 12089 | 2084 |
| PTP1_10_6_16 | 15979 | 2035 |
| PTP1_30_5_15_20 | 6361 | 2052 |
| PTP1_30_6_15_20 | 9114 | 1769 |
| PTP1_40_6_16_20 | 7491 | 1971 |

## DISCUSSION

In this paper, we showed that classical approach for oligo design is not always adapted to complex biological models. The critical point is cross-hybridization that could introduce misinterpretation of biological results. By conventional probe design we were able to determine specific 50-mers oligonucleotides for only approximately 40% of the CDSs in our biological model *Encephalitozoon cuniculi*. We developed a new probe design strategy (GoArray) based on determination of two specific subsequences. In this context, we are able to design specific probes for each CDS of the genome, except perfect duplications. Experimental tests have demonstrated that oligonucleotides designed with our new method can hybridize their targets with a high specificity and sensitivity.

The previous sections showed that oligo characteristics like loop size or linker size seem to be important for the quality of hybridization between new design probe and its target. Future biological experiments will demonstrate the influence of oligonucleotide characteristics on hybridization.

In addition to the originality of the oligo design implemented in GoArray (figure 4), we have paid attention to the quality of the software design following the Object Management Group recommendation (use and extension of the MAGE Object-Model) (Hill *et al.*, 2002). However, it is difficult to compare the results of oligo design software. Indeed, for the same gene, two programs can provide different oligos, but the latter can be completely satisfactory and compatible with the criteria provided by the user. Only two software considering exactly the same criteria for their research and proposing the same user options should provide the same results. If we can provide the same results then the efficiency of the software in computing time can be an additional comparison element. Our prototype, will therefore be rewritten in C++, we have already achieved some testing of distributed implementation (Rimour *et al.*, 2003) and the development of a Web-based oligonucleotide database is envisaged.

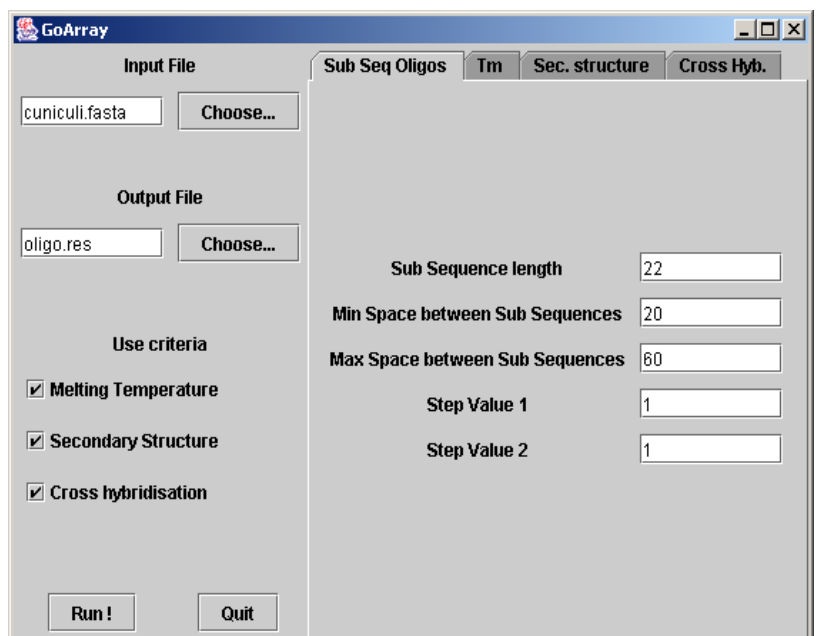

**Fig. 5.** Screenshot of GoArray interface

## REFERENCES

Altschul, S.F., Madden, T.L., Schaffer, A.A., Zhang, J., Zhang, Z., Miller, W. and Lipman, D.J. (1997) Gapped BLAST and PSI-BLAST: a new generation of protein database search programs, *Nucleic Acids Res.*, **25**, 3389-3402.
Bernal, A., Ear, U. and Kyrpides, N. (2001) Genomes OnLine Database (GOLD): a monitor of genome projects world-wide, *Nucleic Acids Res.,* **29**, 126-127.

Chen, H. and Sharp, B.M. (2002) Oliz, a suite of Perl scripts that assist in the design of microarrays using 50mer oligonucleotides from the 3' untranslated region, *BMC Bioinformatics,* **3**:27.

DeRisi, J.L., Iyer, V.R. and Brown, P.O. (1997) Exploring the metabolic and genetic control of gene expression on a genomic scale, *Science,* **278**, 680-286.

Hill D., Rimour S. and Peyret P. (2002) MAGE compliant Object Model for Oligonucleotide Design in Microarray experiments: Application to the eukaryotic parasite Encephalitozoon cuniculi, In *Proceedings of Objects in Bio- & Chem-Informatics*, OMG Object Management Group, Washington, DC (USA), pp 39-44.

Hughes, T.R., Mao, M., Jones, A.R., Burchard, J., Marton, M.J., Shannon, K.W., Lefkowitz, S.M., Ziman, M., Schelter, J.M., Meyer, M.R., Kobayashi, S., Davis, C., Dai, H., He, Y.D., Stephaniants, S.B., Cavet, G., Walker, W.L., West, A., Coffey, E., Shoemaker, D.D., Stoughton, R., Blanchard, A.P., Friend, S.H. and Linsley, P.S. (2001) Expression profiling using microarrays fabricated by an ink-jet oligonucleotide synthesizer, *Nat. Biotechnol.,* **19**, 342-347.

Kane, M.D., Jatkoe, T.A., Stumpf, C.R., Lu, J., Thomas, J.D. and Madore, S.J. (2000) Assessment of the sensitivity and specificity of oligonucleotide (50mer) microarrays, *Nucleic Acids Res.*, **28**, 4552-4557.

Katinka, M.D., Duprat, S., Cornillot, E., Metenier, G., Thomarat, F., Prensier, G., Barbe, V., Peyretaillade, E., Brottier, P., Wincker, P., Delbac, F., El Alaoui, H., Peyret, P., Saurin, W., Gouy, M., Weissenbach, J. and Vivares, C.P. (2001) Genome sequence and gene compaction of the eukaryote parasite Encephalitozoon cuniculi, *Nature,* **414**, 450-453.

Kurata K., Breton V., Saguez C. and Dine G. (2003) Evaluation of Unique Sequences on the European Data Grid., In *proceedings of Asia Pacific Bioinformatics Conference* ( APBC 2003), pp 43-52.

Li, F. and Stormo G.D. (2001) Selection of optimal DNA oligos for gene expression arrays, *Bioinformatics,* **17**, 1067-1076.

Lipshutz, R.J., Fodor, S.P., Gingeras, T.R. and Lockhart, D.J. (1999) High density synthetic oligonucleotide arrays, *Nat. Genet.,* **21**, 20-24.

Lockhart, D.J. and Barlow, C. (2001) Expressing what's on your mind: DNA arrays and the brain, *Nat. Rev. Neurosci,* **2**, 63-68.

Lockhart, D.J., Dong, H., Byrne, M.C., Follettie, M.T., Gallo, M.V., Chee, M.S., Mittmann, M., Wang, C., Kobayashi, M., Horton, H. and Brown, E.L. (1996) Expression monitoring by hybridization to high-density oligonucleotide arrays, *Nat. Biotechnol.*, **14**, 1675-1680.

Nielsen, H.B., Wernersson, R. and Knudsen, S. (2003) Design of oligonucleotides for microarrays and perspectives for design of multi-transcriptome arrays, *Nucleic Acids Res.*, **31**, 3491-3496.

Object Management Group (2003) Gene Expression, v1.1, OMG Final Adopted Specification, OMG document formal/03-10-01.

Raddatz, G., Dehio, M., Meyer, T.F. and Dehio, C. (2001) PrimeArray: genome-scale primer design for DNA-microarray construction, *Bioinformatics*, **17**, 98-99.

Rahmann, S. (2003) Fast Large Scale Oligonucleotide Selection Using the Longest Common Factor Approach, *Journal of Bioinformatics and Computational Biology*, **1**, 343-361.

Relogio, A., Schwager, C., Richter, A., Ansorge, W. and Valcarcel, J. (2002) Optimization of oligonucleotide-based DNA microarrays, *Nucleic Acids Res.*, **30**, e51.

Reymond, N., Charles, H., Duret, L., Calevro, F., Beslon, G. and Fayard, J.M. (2004) ROSO: optimizing oligonucleotide probes for microarrays, *Bioinformatics*, **20**, 271-273.

Rimour S., Hill D. and Peyret P. (2003) Using distributed computing for the design of oligonucleotides in micro-arrays experiments, In *Proceedings of the first Healthgrid Conference*, Lyon, France, pp. 88-96.

Rouillard, J.M., Herbert, C.J. and Zuker, M. (2002) OligoArray: genome-scale oligonucleotide design for microarrays, *Bioinformatics,* **18**, 486-487.

Rouillard, J.M., Zuker, M. and Gulari, E. (2003) OligoArray 2.0: design of oligonucleotide probes for DNA microarrays using a thermodynamic approach, *Nucleic Acids Res.*, **31**, 3057-3062.

Sambrook, J. and Russell, D.W. (2000) Molecular Cloning: A Laboratory Manual. Third edition. Cold Spring Harbor Laboratory Press.

SantaLucia John Jr. (1998) A unified view of polymer, dumbbell, and oligonucleotide DNA nearest-neighbor thermodynamics, *Proc Natl Acad Sci USA,* **95**, 1460-1465.

Schena, M., Shalon, D., Davis, R.W. and Brown, P.O. (1995) Quantitative monitoring of gene expression patterns with a complementary DNA microarray, *Science*, **270**, 467-470.

Talla E, Tekaia F, Brino L. and Dujon B. (2003) A novel design of whole-genome microarray probes for Saccharomyces cerevisiae which minimizes cross-hybridization, *BMC Genomics*, **4**, 38.

Wang, X. and Seed, B. (2003) Selection of oligonucleotide probes for protein coding sequences, *Bioinformatics*, **19**, 796-802.

Zuker, M., Mathews, D.H. and Turner, D.H. (1999) Algorithms and thermodynamics for RNA scondary structure prediction: A Practical Guide, NATO ASI Series, Kluwer, Dordrecht.